%% file: paper.tex
\documentclass{article}
\usepackage{amsmath,amssymb}
\usepackage{color}
\usepackage{epsfig}
\usepackage{subfig}
\usepackage{amsthm}

\input{defines}

\begin{document}
\title{Unbiased Matrix Rounding}
\date{}
\author{Benjamin Doerr\footnotemark[1] \and Tobias Friedrich\footnotemark[1] \and Christian Klein\footnotemark[1] \and Ralf Osbild\thanks{Max-Planck-Institut f\"ur Informatik, Saarbr\"ucken, Germany}}
\maketitle
\CHANGES{ \textcolor{blue}{ \centering Last Change: \input{tmp.latest}} }
\input{abstract}
\section{Introduction}
\label{sec:introduction}

\input{introduction}
\subsection{Knuth's Two-way Rounding}
\label{sec:knuth}

\input{knuth}

\section{Preliminaries}
\label{sec:extension}
\label{sec:preliminaries}

\input{preliminaries}

\section{Bitwise Rounding}
\label{sec:bitwise}

\input{bitwise} %
\subsection{Rounding Half-Integral Matrices}
\input{onehalf}

\subsection{Final Result}
\label{sec:bitwiseresult}

\input{bitwiseresult}

\section{Unbiased Rounding}
\label{sec:unbiased}

\input{unbiased}

\bibliography{paper}
\CHANGES{\newpage Last changes done:\input{tmp.changes}}

\end{document}

%% file: defines.tex
\newcommand{\N}{{\mathbb{N}}}
\newcommand{\R}{{\mathbb{R}}}
\newcommand{\Z}{{\mathbb{Z}}}
\newcommand{\tx}{\ensuremath{\tilde{x}}}
\newcommand{\OH}{\tfrac{1}{2}} 
\newcommand{\GX}{{\cal G}_X} 
\newcommand{\rr}{\ensuremath{\approx}}

\newtheorem{theorem}{Theorem}
\newtheorem{lemma}[theorem]{Lemma}

\newtheorem{definition}[theorem]{Definition}

\newcommand{\tox}[1]{\ensuremath{[1..#1]}}
\newcommand{\ton}[1]{\ensuremath{#1\in\tox{n}}}

\newcommand{\tom}[1]{\ensuremath{#1\in\tox{m}}}

\newcommand{\NOTE}[2]{}
\newcommand{\CHANGES}[1]{}

%% file: abstract.tex
\begin{abstract}
\noindent
We show several ways to round a real matrix to an integer one such 
that the rounding errors in all rows and columns as well as the whole
matrix are less than one. This is a classical problem with applications in
many fields, in particular, statistics.

We improve earlier solutions of different authors in two ways. For
rounding matrices of size $m \times n$, we reduce the runtime from $O( (m n)^2 )$ to $O(m n \log(m n))$.
Second, our roundings also have a rounding error of less than one in all
initial intervals of rows and columns. Consequently, arbitrary intervals
have an error of at most two. This is particularly useful in the
statistics application of controlled rounding.

The same result can be obtained via (dependent) randomized rounding. This
has the additional advantage that the rounding is unbiased, that is, for
all entries $y_{ij}$ of our rounding, we have $E(y_{ij}) = x_{ij}$, where
$x_{ij}$ is the corresponding entry of the input matrix.
\end{abstract}

%% file: introduction.tex
In this paper, we analyze a rounding problem with strong connections to statistics, but also to different areas in discrete mathematics, computer science, and operations research.
We show how to round a matrix to an integer one such that rounding errors in intervals of rows and columns are small.

Let $m, n$ be positive integers.  For some set $S$, we write
$S^{m\times n}$ to denote the set of $m\times n$ matrices with entries
in $S$.  For real numbers $a, b$ let $[a..b] := \{ z \in \Z \mid a\le
z\le b\}$. We show the following.

\begin{theorem}\label{tbinary}
  For all $X \in [0, 1)^{m \times n}$ a rounding $Y \in \{0,1\}^{m
    \times n}$ such that
  \begin{eqnarray*}
    \forall b \in [1..n], \; i \in [1..m] &:& \bigg|\sum_{j = 1}^b (x_{ij} - y_{ij})\bigg| < 1,\\
    \forall b \in [1..m], \; j \in [1..n] &:& \bigg|\sum_{i = 1}^b (x_{ij} - y_{ij})\bigg| < 1,\\
                                          & & \bigg|\sum_{i = 1}^m\sum_{j = 1}^n (x_{ij} - y_{ij})\bigg| < 1
  \end{eqnarray*}
  can be computed in time $O(m n \log(m n))$.   
\end{theorem}\noindent
This result extends the famous rounding lemma of
Baranyai~\cite{baranyai} and several results on controlled rounding in
statistics by Bacharach~\cite{bacharach66} and Causey, Cox and Ernst~\cite{Causey}.

\subsection{Baranyai's Rounding Lemma and Applications in Statistics}
\label{sec:rnd}%
Baranyai~\cite{baranyai} used a weaker version of Theorem~\ref{tbinary} to obtain his
well-known results on coloring and partitioning complete uniform
hypergraphs. He showed that any matrix can be rounded such that
the errors in all rows, all columns and the whole matrix are less than
one. He used a formulation as flow problem to prove this
statement. This yields an inferior runtime than the bound in
Theorem~\ref{tbinary}. However, algorithmic issues were not his focus.

In statistics, Baranyai's result was independently obtained by
Bacharach~\cite{bacharach66} (in a slightly weaker form) and again independently by Causey, Cox
and Ernst~\cite{Causey}.
There are two statistical applications for such rounding results.
Note first that instead of rounding to integers, our result also applies to rounding to multiples of any other base (e.g., multiples of 10).
Such a rounding can be used to improve the readability of data tables.

The main reason, however, to apply such a rounding procedure is confidentiality protection.
Frequency counts that directly or indirectly disclose small counts may permit the identification of individual respondents.
There are various methods to prevent this~\cite{lns155}, one of which is \emph{controlled rounding}~\cite{ControlledRounding}.
Here, one tries to round an $(m+1) \times (n+1)$-table $\tilde{X}$ given by
\[
\begin{array}{c|c}
  \left(x_{i j}\right)_{\genfrac{}{}{0pt}{}{i=1\ldots m}{j=1 \ldots n} }     & \left(\sum_{j=1}^{n} x_{i j}\right)_{i=1\ldots m} \\\\\hline\\
  \left(\sum_{i=1}^{m} x_{i j}\right)_{j=1 \ldots n}           & \sum_{i=1}^{m}\sum_{j=1}^{n} x_{i j} \\
\end{array}\noindent
\]
to an $(m+1) \times (n+1)$-table $\tilde{Y}$ such that additivity is preserved, i.e., the last row and column of $\tilde{Y}$ contain the associated totals of $\tilde{Y}$.
In our setting we round the $m \times n$-matrix $X$ defined by the $m n$ inner cells of the table $\tilde{X}$ to obtain a controlled rounding.

The additivity in the rounded table allows to derive information on the row and column totals of the original table.
In contrast to other rounding algorithms, our result also permits to retrieve further reliable information from the rounded matrix,
namely on the sums of consecutive elements in rows or columns.
Such queries may occur if there is a linear ordering on statistical attributes. Here an example. Let $x_{ij}$ be the number of people in
country $i$ that are $j$ years old. Say $Y$ is such that $\frac{1}{1000} Y$ is a rounding of $\frac{1}{1000} X$ as in Theorem~\ref{tbinary}.
Now $\sum_{j = 20}^{40} y_{ij}$ is the number of people in country $i$ that are between $20$ and $40$ years old, apart from an error of less than $2000$.
Note that such guarantees are not provided by the results of Baranyai~\cite{baranyai}, Bacharach~\cite{bacharach66}, and Causey, Cox and Ernst~\cite{Causey}. 

\subsection{Unbiased Rounding}
In Section~\ref{sec:unbiased}, we present a randomized algorithm computing roundings as in  Theorem~\ref{tbinary}.
It has the additional property that each matrix entry is rounded up with probability equal to its fractional value.
This is known as randomized rounding \cite{ragh} in computer science and as unbiased controlled rounding \cite{unbiasedCR,fellegi75} in statistics.
Here, a controlled rounding is computed such that the expected values of each table entry (including the totals) equals its fractional value in the original table.

To state our result more precisely, we introduce the following notation.
For $x\in\R$ write $\lfloor x \rfloor := \max\{z\in\Z \mid z \le r\} , \lceil x \rceil := \min\{z\in\Z \mid z \ge r\}$ and $\{x\} := x - \lfloor x \rfloor$.
\begin{definition}
  Let $x\in\R$.
  A random variable $y$ is called \emph{randomized rounding} of $x$, denoted $y \rr x$, if $\Pr(y=\lfloor x \rfloor +1) = \{x\}$ and $\Pr(y=\lfloor x \rfloor ) = 1-\{x\}$.
  For a matrix $X\in\R^{m \times n}$, we call an ${m \times n}$ matrix-valued random variable $Y$ randomized rounding of $X$ if $y_{i j} \rr x_{i j}$ for all $\tom{i},\ton{j}$.
\end{definition}\noindent
We then get the following randomized version of Theorem~\ref{tbinary}.
\begin{theorem}\label{tunbi}
  Let $X \in [0, 1)^{m \times n}$ be a matrix having entries of binary length at most $\ell$.
  Then a randomized rounding $Y$ fulfilling the additional constraints that
  \begin{eqnarray*}
    \forall b \in [1..n], \; i \in [1..m] &:& \sum_{j = 1}^b x_{ij} \rr \sum_{j = 1}^b y_{ij},\\
    \forall b \in [1..m], \; j \in [1..n] &:& \sum_{i = 1}^b x_{ij} \rr \sum_{i = 1}^b y_{ij},\\
                                          & & \sum_{i = 1}^m\sum_{j = 1}^n x_{ij} \rr \sum_{i = 1}^m\sum_{j = 1}^n y_{ij}
  \end{eqnarray*}
  can be computed in time $O(m n \ell)$.
\end{theorem}\noindent
For a matrix with arbitrary entries $x_{i j}:=\sum_{d=1}^{\ell} x_{i j}^{(d)} 2^{-d} + x_{i j}'$ where $x_{i j}'<2^{-\ell}$ and $x_{i j}^{(d)}\in\{0,1\}$ for $\tom{i},\ton{j}, d\in\tox{\ell}$,
we may use the $\ell$ highest bits to get an approximate randomized rounding.
If (before doing so) we round the remaining part $x_{i j}'$ of each entry to $2^{-\ell}$ with probability $2^{\ell}x_{i j}'$ and to $0$ otherwise, we still have that $Y \rr X$, but we introduce an additional error of at most $2^{-\ell}m n$ in the constraints of Theorem~\ref{tunbi}.

\subsection{Other  Applications}
One of the most basic rounding results states that any sequence $x_1,
\ldots, x_n$ of numbers can be rounded to an integer one $y_1, \ldots,
y_n$ such that the rounding errors $|\sum_{j = a}^b (x_j -
y_j)|$ are less than one for all $a, b \in [1..n]$. Such roundings can
be computed efficiently in linear time by a one-pass algorithm
resembling Kadane's scanning algorithm (described in Bentley's
Programming Pearls~\cite{bentley}).
Extensions in different directions
have been obtained in~\cite{ichlindtu,ichsequences,knuth,halficalp,10l}.
This rounding problem has found a number of applications, among
others in image processing~\cite{asanoieice,halfdagstuhl}.

Theorem~\ref{tbinary} extends this result to two-dimensional sequences.
Here the rounding error in arbitrary intervals of a row or column is less than two.
In \cite{DFKO05} a lower bound of 1.5 is shown for this problem.
Thus an error of less than one as in the one-dimensional case cannot be achieved.

Rounding a matrix while considering the errors in column sums and partial row sums also arises in scheduling~\cite{brauner,monden2,monden1,steineryeomans}.
For this, however, one does not need our result in full generality.
It suffices to use the linear-time one-pass algorithm given in \cite{DFKO05}.
This algorithm rounds a matrix having unit column sums and can be extend to compute a quasi rounding for arbitrary matrices.
While this algorithm keeps the error in all initial row intervals small, for columns only the error over the whole column is considered.

%% file: knuth.tex
In~\cite{knuth}, Knuth showed how to round a sequence of $n$ real numbers $x_i$ to $y_i \in \left\{\lfloor x_i \rfloor, \lceil x_i \rceil \right\}$ such that for two given permutations $\sigma_1, \sigma_2\in S_n$, we have both $|\sum_{i=1}^{k} (x_{\sigma_1(i)}-y_{\sigma_1(i)})|\leq n / (n + 1)$ and $|\sum_{i=1}^{k} (x_{\sigma_2(i)}-y_{\sigma_2(i)})|\leq n / (n + 1)$ for all $k$.
Knuth's proof uses integer flows in a certain network~\cite{fordfulkerson}.
On account of this his worst-case runtime is quadratic.

One application Knuth mentioned in~\cite{knuth} is that of matrix rounding.
For this, simply choose a permutation $\sigma_1$ that enumerates the $x_{i j}$ row by row, and a permutation $\sigma_2$ that enumerates the $x_{i j}$ column by column.
Applying Knuth's algorithm to these permutations gives a rounding with errors smaller than one in all initial row and column intervals.

%% file: preliminaries.tex
In this section, we provide two easy extensions of the result
stated in the introduction. First, we
immediately obtain rounding errors of less than two in arbitrary
intervals in rows and columns. This is supplied by the following lemma.

\begin{lemma}\label{linitial}
  Let $Y$ be a rounding of a matrix $X$ such that the errors
  $|\sum_{j = 1}^b (x_{i j} - y_{i j})|$
  in all initial intervals of rows are  at most
  $d$. Then the errors in arbitrary intervals of rows are at most $2 d$,
  that is, for all $i \in [1..m]$ and all $1 \le a \le b \le n$,
  \[\bigg|\sum_{j = a}^b (x_{i j} - y_{i j}) \bigg| \le 2d.\]
  This also holds for column intervals, i.e., if 
  the errors $|\sum_{i = 1}^b (x_{i j} - y_{i j})|$ in all initial intervals of columns are at most $d'$,
  then the errors $|\sum_{i = a}^b (x_{i j} - y_{i j})|$ in arbitrary intervals of columns are at most $2 d'$.
\end{lemma}

\begin{proof}
  Let $i \in [1..m]$ and $1 \le a \le b \le n$. Then 
  \begin{eqnarray*}
    \bigg|\sum_{j =
    a}^b (x_{i j} - y_{i j}) \bigg| &=& \bigg|\sum_{j = 1}^b (x_{i j} -
  y_{i j}) - \sum_{j = 1}^{a-1} (x_{i j} - y_{i j})\bigg| \\ &\le&
  \bigg|\sum_{j = 1}^b (x_{i j} - y_{i j})\bigg| + \bigg|\sum_{j =
    1}^{a-1} (x_{i j} - y_{i j})\bigg|\ \le\ 2d.
  \end{eqnarray*}
\end{proof}\noindent
From now on, we will only consider matrices having integral row and column sums.
This is justified by the following lemma.
\begin{lemma}\label{lint}
  Assume that for any $X \in \R^{m \times n}$
  with integral column and row sums
  a rounding $Y \in \Z^{m \times n}$ such that
  \begin{eqnarray}
     \forall b \in [1..n], \; i \in [1..m] &:& \bigg|\sum_{j = 1}^b (x_{i j} - y_{i j})\bigg| < 1, \label{eq:lem1a}\\
     \forall b \in [1..m], \; j \in [1..n] &:& \bigg|\sum_{i = 1}^b (x_{i j} - y_{i j})\bigg| < 1  \label{eq:lem1b}
  \end{eqnarray}
  can be computed in time T(m,n).
  Then for all $\tilde X \in \R^{m \times n}$
  with arbitrary column and row sums
  a rounding $\tilde Y \in \Z^{m \times n}$ satisfying \eqref{eq:lem1a}, \eqref{eq:lem1b} and
  \begin{eqnarray}
     \bigg|\sum_{i = 1}^m \sum_{j = 1}^n (x_{i j} - y_{i j}) \bigg| < 1 \label{eq:lem1c}
  \end{eqnarray}
  can be computed in time $T(m+1,n+1)+O(m n)$.
\end{lemma}

\begin{proof}
  Given an arbitrary matrix $\tilde X \in \R^{m \times n}$, we add an extra row taking what is
  missing towards integral column sums and add an extra column taking
  what is missing towards integral row sums.
  Hence, let $X \in \R^{(m+1) \times (n+1)}$ be such that
  \begin{eqnarray*}
      x_{i j}     = \tilde x_{i j} &&\text{ for all } i \in [1..m], \; j \in [1..n],\\
      x_{m+1,j}   = \bigg\lceil \sum_{i = 1}^m \tilde x_{i j} \bigg\rceil - \sum_{i = 1}^m \tilde x_{i j} &&\text{ for all } j \in [1..n], \\
      x_{i,n+1}   = \bigg\lceil \sum_{j = 1}^n \tilde x_{i j} \bigg\rceil - \sum_{j = 1}^n \tilde x_{i j} &&\text{ for all } i \in [1..m], \\
      x_{m+1,n+1} = \bigg\lceil \sum_{i = 1}^m \tilde x_{i,n+1} \bigg\rceil - \sum_{i = 1}^m \tilde x_{i,n+1} &=& \bigg\lceil \sum_{j = 1}^n \tilde x_{m+1,j} \bigg\rceil - \sum_{j = 1}^n \tilde x_{m+1,j}.
  \end{eqnarray*}
  Clearly, $X$ has integral row and column sums.
  Therefore it can be rounded to $Y\in\Z^{(m+1) \times (n+1)}$ satisfying \eqref{eq:lem1a} and \eqref{eq:lem1b} in time $T(m+1,n+1)$.

  For \eqref{eq:lem1c}, observe that if a row (resp. column) sum is integral, the rounding error in the row (resp. column) is $0$.
  Then the rounding error in the whole matrix is also $0$, if all row and column sums are integral.
  Using this and the triangle inequality, we get inequality \eqref{eq:lem1c} as follows.
  \begin{eqnarray*}
    \bigg|\sum_{i = 1}^m \sum_{j = 1}^n (x_{i j} - y_{i j}) \bigg| & = &
    \bigg|\sum_{i = 1}^{m+1} \sum_{j = 1}^{n+1} (x_{i j} - y_{i j}) - \sum_{i = 1}^{m+1} (x_{i,n+1} - y_{i,n+1})\\
      &     & -\sum_{j = 1}^{n+1} (x_{m+1,j} - y_{m+1,j}) + (x_{m+1,n+1} - y_{m+1,n+1})\bigg| \\
      & \le & 0 + 0 + 0 + |x_{m+1,n+1} - y_{m+1,n+1}| < 1.
  \end{eqnarray*}
  By setting $\tilde y_{i j} = y_{i j}$ for all $i \in [1..m]$ and $j \in [1..n]$, we obtain the desired rounding
  $\tilde Y \in \Z^{m \times n}$.
\end{proof}\noindent

%% file: bitwise.tex
In this section, we present an alternative approach
which will lead to a superior runtime. It uses a
classical result on rounding problems, namely, that the problem of
rounding arbitrary numbers can be reduced to the one of rounding
half-integral numbers.
For $X\in\{0,\OH\}^{m\times n}$, our rounding problem
turns out to be much simpler.
In fact, it can be solved in linear time.

\subsection{The Binary Rounding Method}

The following rounding method was introduced by Beck and Spencer~\cite{bslong} in 1984.
They used it to prove the existence of two-colorings of $\N$ having small discrepancy in all arithmetic progressions of arbitrary length and bounded difference.

Given arbitrary numbers that have to
be rounded, they use their binary expansion and (assuming all of
them to be finite) round `digit by digit'. To do the latter, they only
need to understand the corresponding rounding problem for
half-integral numbers.
That is, an $\ell$-bit number $x=x'+\OH x'',
x'\in\{0,\OH\}$ can be recursively rounded by rounding
the $(\ell-1)$-bit number $x''$ to $y''\in\{0,1\}$ and
then rounding $x'+\OH y''\in\{0,\OH,1\}$ to $y\in\{0,1\}$.
The resulting rounding errors are at most twice the ones incurred by the half-integral roundings.

If some numbers do not have a finite binary expansion,
one can use a sufficiently large finite length approximation.
To get rid of additional errors caused by this,
we invoke a slight refinement of the binary rounding method.
In~\cite{ichlind} it was proven that the extra factor of two
can be reduced to an extra factor of $2(1-\frac{1}{2r})$,
where $r$ is the number of rounding errors we want to keep small.

In our setting, the number of rounding errors is the number of
all initial row and column intervals, i.e., $r=2m n$.
In summary, we have the following.

\begin{lemma}\label{lbinary}
  Assume that for any $X \in \{0, \frac 12\}^{m \times n}$ a rounding
  $Y \in \{0,1\}^{m \times n}$ can be computed in time $T$ that
  satisfies
	\begin{eqnarray*}
	&&\forall b \in [1..n], \; i \in [1..m] : \bigg|\sum_{j = 1}^b (x_{i j} - y_{i j})\bigg| \le D,\\
	&&\forall b \in [1..m], \; j \in [1..n] : \bigg|\sum_{i = 1}^b (x_{i j} - y_{i j})\bigg| \le D.
	\end{eqnarray*}
Then for all $\ell \in \N$ and $X \in [0, 1)^{m \times n}$ a rounding $Y \in \{0,1\}^{m \times n}$ such that 
	\begin{eqnarray*}
	&&\forall b \in [1..n], \; i \in [1..m] : \bigg|\sum_{j = 1}^b (x_{i j} - y_{i j})\bigg| \le 2(1-\tfrac{1}{4m n})D + 2^{-\ell}b,\\
	&&\forall b \in [1..m], \; j \in [1..n] : \bigg|\sum_{i = 1}^b (x_{i j} - y_{i j})\bigg| \le 2(1-\tfrac{1}{4m n})D + 2^{-\ell}b	 
	\end{eqnarray*}
	can be computed in time $O(\ell\, T)$. 
\end{lemma}

%% file: onehalf.tex
It remains to show how to solve the rounding problem for half-integral matrices.
Based on Lemma~\ref{lint},
 we can assume integrality of row and column sums.

Here is an outline of our approach.
For each row and column,
 we consider the sequence of its $\OH$--entries
 and partition them into disjoint pairs of neighbors.
From the two $\OH$s forming such a pair, exactly one is rounded to $1$
and the other to $0$.
Thus, if such a pair is contained in an initial interval,
 it does not contribute to the rounding error.

\begin{figure}[tbh]
\centering
\subfloat[]{
    \includegraphics[bb=70pt 403pt 515pt 689pt,width=.47\textwidth,clip]{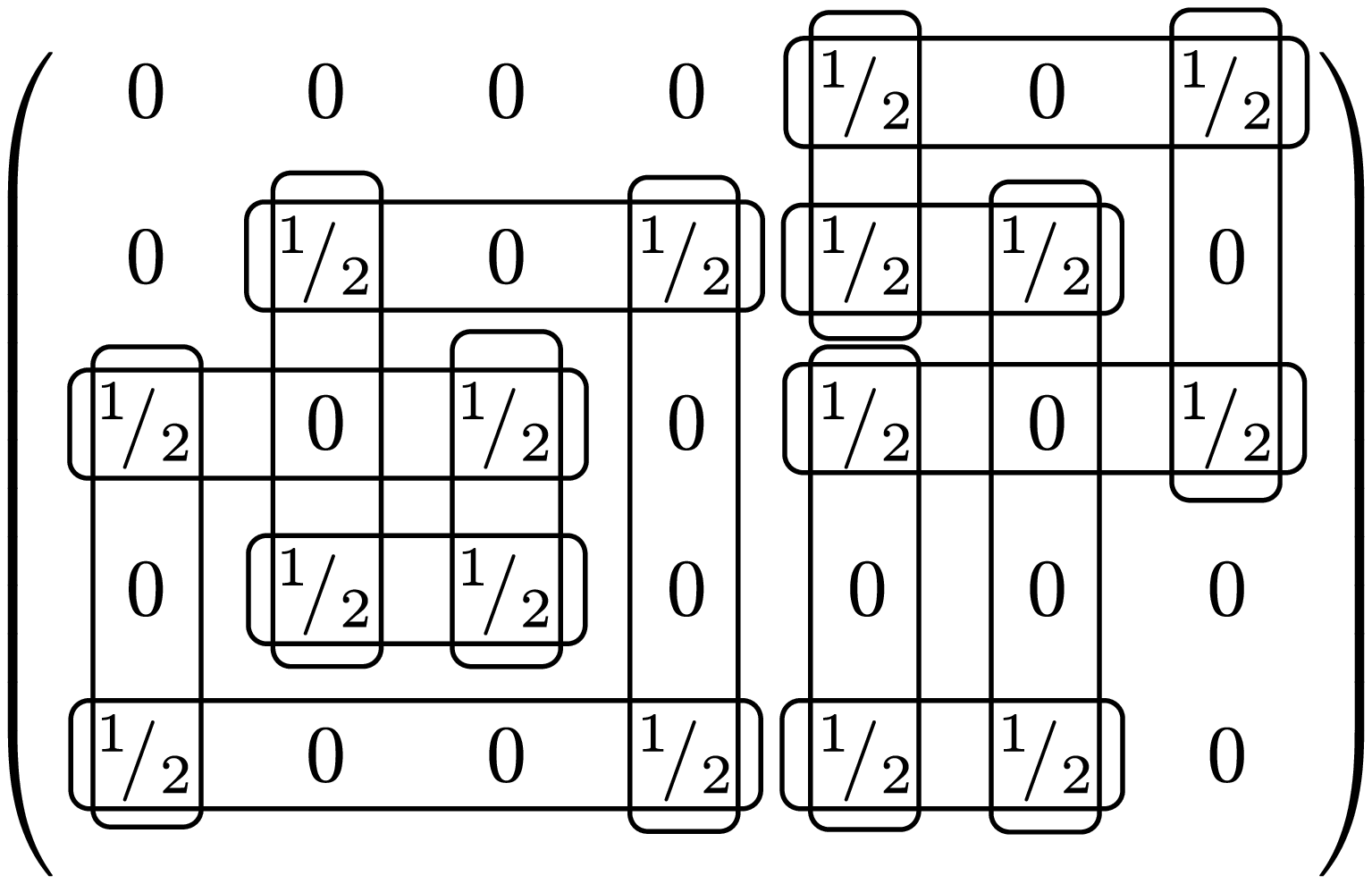}
    \label{subfig:matrix}
}
\subfloat[]{
    \includegraphics[bb=70pt 403pt 515pt 689pt,width=.47\textwidth,clip]{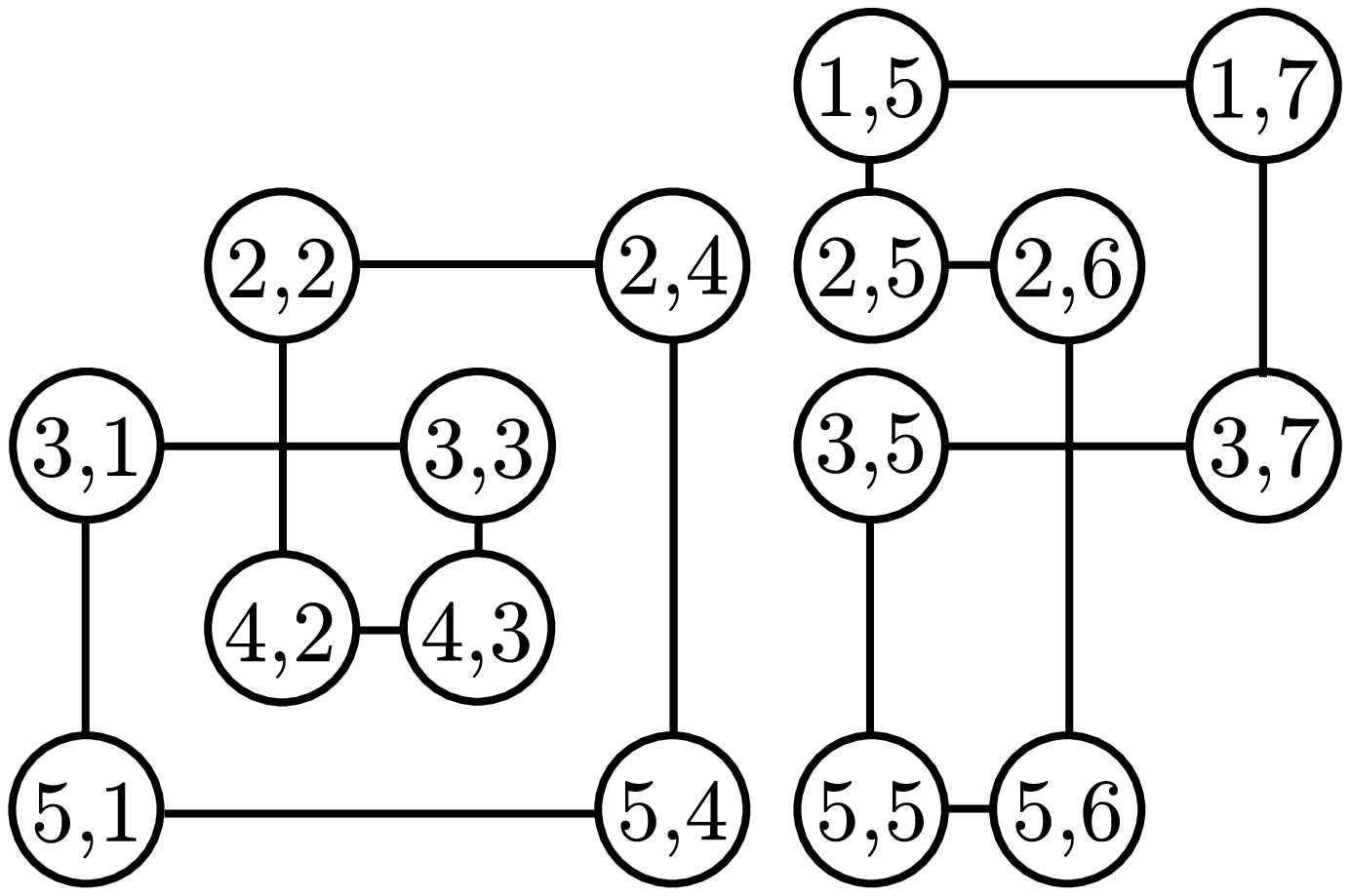}
    \label{subfig:graph}
}
\caption{Example for the construction of an auxiliary graph.
\protect\subref{subfig:matrix} Input matrix $X$ with its row and column pairs.
\protect\subref{subfig:graph} Auxiliary graph $\GX$. Vertices are labeled with matrix indices
     and edges connect vertices of row and column pairs.
     $\GX$ is a disjoint union of even cycles.
}
\label{fig:example}
\end{figure}

To make the idea precise,
 assume some row contains exactly $2K$ entries of value $\OH$.
We call the $(2k-1)$--th and $(2k)$--th $\OH$--entry of this row
 a {\em row pair}, for all $1\le k\le K$.
The $\OH$s of a row pair are mutually referred to as {\em row neighbors}.
Similarly, we define {\em column pairs} and {\em column neighbors}.
Figure~\ref{fig:example}\subref{subfig:matrix}
 shows a half-integral matrix together with
 row and column pairs marked by boxes.
Since each $\OH$ belongs to a row pair {\em and} a column pair,
 the task of rounding is non-trivial.

Our solution makes use of an auxiliary graph $\GX$
 which contains the necessary information about row and column neighbors.
Each $\OH$--entry is represented by a vertex
 that is labeled with the corresponding matrix indices.
Each pair is represented by an edge connecting the vertices
 that correspond to the paired $\OH$s.
Figure~\ref{fig:example}\subref{subfig:graph} shows the auxiliary graph
 that belongs to the matrix of Figure~\ref{fig:example}\subref{subfig:matrix}.

We collect some properties of this auxiliary graph.

\begin{lemma}\label{lem:cycleprop}
  Let $X \in \{0,\tfrac{1}{2}\}^{m \times n}$ be a matrix with integral row and column sums.
   \begin{enumerate}
     \item[(a)] Every vertex of $\GX$ has degree 2.
     \item[(b)] $\GX$ is a disjoint union of even cycles.
     \item[(c)] $\GX$ is bipartite.
   \end{enumerate}
\end{lemma}
\begin{proof}
(a)
Because of the integrality of the row and column sums,
 the number of $\OH$--entries in each row and column is even.
Hence each $\OH$--entry has a row and a column neighbor.
In consequence, each vertex is incident with exactly two edges.
(b)
The edge sequence of a path in $\GX$
 corresponds to an alternating sequence of row and column pairs.
Therefore any cycle in $\GX$ consists of an even number of edges.
Since each vertex has degree two, $\GX$ is a disjoint union of cycles.
(c)
Clearly, every even cycle is bipartite.
\end{proof}\noindent
With this result, we are able to find the desired roundings.
\begin{lemma}\label{lem:bipartite}
Let $X \in \{0,\tfrac{1}{2}\}^{m \times n}$ and
 let $V_0 \dot\cup V_1$ be a bipartition of $\GX$.
Define $Y=(y_{ij})\in \{0,1\}^{m\times n}$ by
  \begin{eqnarray*}
    y_{ij}=\begin{cases}
      0, & \text{if }  x_{ij}=0\\
      0, & \text{if }  x_{ij}=\OH \text{ and } (i,j)\in V_0\\
      1, & \text{if }  x_{ij}=\OH \text{ and } (i,j)\in V_1.
    \end{cases}
  \end{eqnarray*}
Then $Y$ has the property that
  \begin{eqnarray}
    \label{eqn:initrow}
    \forall b \in [1..n],\; i \in [1..m] &:& \bigg|\sum_{j = 1}^b (x_{ij} - y_{ij})\bigg| \le \tfrac{1}{2},\\
    \label{eqn:initcol}
    \forall b \in [1..m],\; j \in [1..n] &:& \bigg|\sum_{i = 1}^b (x_{ij} - y_{ij})\bigg| \le \tfrac{1}{2}.
  \end{eqnarray}
\end{lemma}
\begin{proof}
 Because $0$s of $X$ are maintained in $Y$,
  it suffices to consider $\OH$--entries
  to determine the rounding error in initial intervals.
 Since the rounded values for the $(2k-1)$-th and $(2k)$-th $\OH$--entry
  sum up to 1 by construction,
  there is no error in initial intervals that contain an {even} number
  of $\OH$s, and an error of $\OH$ if they contain an {odd} number of $\OH$s.
\end{proof}\noindent
After these considerations,
we are able to present an algorithm that solves the problem in two steps:
 first we compute the auxiliary graph
 and afterwards the output matrix.
To construct $\GX$, we transform the input matrix $X$
 column by column from left to right.
Of course, generating the labeled vertices is trivial.
The column neighbors are detected just by numbering the $\OH$--entries
 within a column from top to bottom.
When there are $2k$ such entries,
 we insert an edge between the vertices with number $2i-1$ and $2i$
 with $1\le i \le k$.
The strategy to detect row neighbors is the same but
 we need more information.
Therefore we store for each row the parity of its $\OH$--entries so far
 and, if the parity is odd,
 further a pointer to the last occurrence of $\OH$ in this row.
Then, if the current $\OH$ is an even occurrence,
 we have a pointer to the preceding $\OH$,
 and are able to insert an edge between the corresponding vertices in $\GX$.

The output matrix $Y$ can be computed from $X$ as follows.
Every $0$ in $X$ is kept and every $\OH$--sequence that corresponds to
 a cycle in $\GX$ is substituted by
 an alternating $0$--$1$--sequence.
By Lemma~\ref{lem:cycleprop}, this is always possible.
It does not matter which of the two alternating $0$--$1$ sequences we choose.

The graph $\GX$ can be realized with adjacency lists
 (the vertex degree is always 2).
The additional information per row can be realized by a
 simple pointer--array of length $m$
 (a special nil--value indicates even parity).

Since the runtime of each step is bounded
 by the size of the input matrix,
 the entire algorithm takes time $O(m n)$.
In addition to the constant amount of space we need for each of the $m$ rows,
 we store all $k$ entries of value $\tfrac{1}{2}$ in the auxiliary graph.
This leads to a total space consumption of $O(m+k)$.
Summarizing the above, we obtain the following lemma.
\begin{lemma}\label{lhalfintegral}
  Let $X \in \{0,\tfrac{1}{2}\}^{m \times n}$.
  Then a rounding $Y\in\{0,1\}^{m\times n}$
  satisfying the inequalities~(\ref{eqn:initrow}) and (\ref{eqn:initcol})
  can be computed in time $O(m n)$.
\end{lemma}

%% file: bitwiseresult.tex
By combining Lemma~\ref{lbinary} and~\ref{lhalfintegral}, we obtain
the following result.

\begin{theorem}
  For all $\ell \in \N$ and $X \in [0, 1)^{m \times n}$ a rounding $Y \in \{0,1\}^{m \times n}$ such that 
\begin{eqnarray*}
	&&\forall b \in [1..n], \; i \in [1..m] : \bigg|\sum_{j = 1}^b (x_{i j} - y_{i j})\bigg| \le 1 - \tfrac{1}{4m n} + 2^{-\ell}b,\\
	&&\forall b \in [1..m], \; j \in [1..n] : \bigg|\sum_{i = 1}^b (x_{i j} - y_{i j})\bigg| \le 1 - \tfrac{1}{4m n} + 2^{-\ell}b	 
\end{eqnarray*}
can be computed in time $O(\ell m n)$. 
\end{theorem}
\noindent
For $\ell > \log_2(4 m n \, \max\{m,n\})$ the above theorem together with Lemma~\ref{lint} yields Theorem~\ref{tbinary} in the introduction.

%% file: unbiased.tex
In this section we give a randomized algorithm that computes a randomized rounding satisfying Theorem~\ref{tunbi}.
First observe, that the $\{0,\OH\}$ case has a very simple randomized solution.
Whenever it has to round a cycle, it chooses one of the two alternating $0$--$1$--sequences for each cycle uniformly at random.
Then, each $x_{i j}=\OH$ is rounded up with probability $\OH$.

Now consider the output of the bitwise rounding algorithm using the randomized rounding algorithm for the half-integral case as subroutine.
We adapt the proofs of \cite{doerrRR06} to show that this algorithm computes an unbiased controlled rounding.
\begin{theorem}\label{rndall}
  Let $X\in[0,1)^{m \times n}$ be a matrix containing entries with binary representation of length at most $\ell$.
  Let $Y$ be a random variable modeling the output of the randomized algorithm.
  Then $Y \rr X$ and
  \begin{eqnarray}
    \forall b \in [1..n], \; i \in [1..m] &:& \sum_{j = 1}^b y_{i j} \rr \sum_{j = 1}^b x_{i j} , \label{eq:lemrnd1}\\
    \forall b \in [1..m], \; j \in [1..n] &:& \sum_{i = 1}^b y_{i j} \rr \sum_{i = 1}^b x_{i j} . \label{eq:lemrnd2}
  \end{eqnarray}
\end{theorem}
\begin{proof}
  We prove $Y \rr X$ by induction.
  For $\ell=1$ it is clear that $\Pr(y_{i j} = 1) = x_{i j}$.
  If $\ell>1$, write $x_{i j} = x_{i j}' + \OH x_{i j}''$, where $x_{i j}'\in\{0,\OH\}$ and $x_{i j}''\in[0,1)$ has bit-length $\ell-1$.
  Let $y_{i j}''$ be the rounding computed for $x_{i j}''$.
  Then $\Pr(y_{i j}'' = 1)=x_{i j}''$ by induction.
  Now the algorithm will round $\tx_{i j} := x_{i j}'+\OH y_{i j}''\in\{0,\OH,1\}$ to $y_{i j}$.
  If $y_{i j}''=1$, then $\tx_{i j}$ will be rounded up with probability $1$ if $x_{i j}'=\OH$ and with probability $\OH$ otherwise.
  If, on the other hand, $y_{i j}''=0$, then $\tx_{i j}$ will be rounded up with probability $x_{i j}'$.
  Thus
  \[ \Pr(y_{i j} = 1) = x_{i j}''(\OH+x_{i j}') + (1-x_{i j}'')x_{i j}' = x_{i j}' + \OH x_{i j}'' = x_{i j} .\]
  \noindent
  To prove equation \eqref{eq:lemrnd1}, observe that $s_y:=\sum_{j = 1}^b y_{i j}$ is a rounding of $s_x:=\sum_{j = 1}^b x_{i j}$ by Lemma~\ref{lbinary}.
We also have $E(s_y) = \sum_{j = 1}^b E(y_{i j}) = s_x$ by linearity of expectation.
  But also $E(s_y) = \Pr(s_y=\lfloor s_x\rfloor) \lfloor s_x \rfloor + \Pr(s_y=\lfloor s_x \rfloor +1) (\lfloor s_x\rfloor +1)$, which is only possible if $s_y \rr s_x$.
  The proof of \eqref{eq:lemrnd2} is analogous.
\end{proof}\noindent

%% file: paper.bbl
\begin{thebibliography}{10}

\bibitem{asanoieice}
T.~Asano.
\newblock Digital halftoning: Algorithm engineering challenges.
\newblock {\em IEICE Trans. on Inf. and Syst.}, E86-D:159--178, 2003.

\bibitem{bacharach66}
M.~Bacharach.
\newblock Matrix rounding problems.
\newblock {\em Management Science (Series A)}, 12:732--742, 1966.

\bibitem{baranyai}
Zs. Baranyai.
\newblock On the factorization of the complete uniform hypergraph.
\newblock In {\em Infinite and finite sets (Colloq., Keszthely, 1973; dedicated
  to P. Erd\H os on his 60th birthday), Vol. I}, pages 91--108. Colloq. Math.
  Soc. J\'an\=os Bolyai, Vol. 10. North-Holland, Amsterdam, 1975.

\bibitem{bslong}
J.~Beck and J.~Spencer.
\newblock Well distributed 2-colorings of integers relative to long arithmetic
  progressions.
\newblock {\em Acta Arithm.}, 43:287--298, 1984.

\bibitem{bentley}
J.~L. Bentley.
\newblock Algorithm design techniques.
\newblock {\em Commun. ACM}, 27:865--871, 1984.

\bibitem{brauner}
N.~Brauner and Y.~Crama.
\newblock The maximum deviation just-in-time scheduling problem.
\newblock {\em Discrete Appl. Math.}, 134:25--50, 2004.

\bibitem{Causey}
B.~D. Causey, L.~H. Cox, and L.~R. Ernst.
\newblock Applications of transportation theory to statistical problems.
\newblock {\em Journal of the American Statistical Association}, 80:903--909,
  1985.

\bibitem{unbiasedCR}
L.~H. Cox.
\newblock A constructive procedure for unbiased controlled rounding.
\newblock {\em Journal of the American Statistical Association}, 82:520--524,
  1987.

\bibitem{ControlledRounding}
L.~H. Cox and L.~R. Ernst.
\newblock Controlled rounding.
\newblock {\em Informes}, 20:423--432, 1982.

\bibitem{ichlind}
B.~Doerr.
\newblock Linear and hereditary discrepancy.
\newblock {\em Combinatorics, Probability and Computing}, 9:349--354, 2000.

\bibitem{ichlindtu}
B.~Doerr.
\newblock Lattice approximation and linear discrepancy of totally unimodular
  matrices.
\newblock In {\em Proceedings of the 12th Annual ACM-SIAM Symposium on Discrete
  Algorithms (SODA)}, pages 119--125, 2001.

\bibitem{ichsequences}
B.~Doerr.
\newblock Global roundings of sequences.
\newblock {\em Information Processing Letters}, 92:113--116, 2004.

\bibitem{doerrRR06}
B.~Doerr.
\newblock Generating randomized roundings with cardinality constraints and
  derandomizations.
\newblock In {\em 23rd Annual Symposium on Theoretical Aspects of Computer
  Science}, 2006.

\bibitem{DFKO05}
B.~Doerr, T.~Friedrich, C.~Klein, and R.~Osbild.
\newblock Rounding of sequences and matrices, with applications.
\newblock In {\em Third Workshop on Approximation and Online Algorithms},
  volume 3879 of {\em Lecture Notes in Computer Science}, pages 96--109.
  Springer, 2006.

\bibitem{fellegi75}
I.~P. Fellegi.
\newblock Controlled random rounding.
\newblock {\em Survey Methodology}, 1:123--133, 1975.

\bibitem{fordfulkerson}
L.~R. Ford, {Jr.,} and D.~R. Fulkerson.
\newblock {\em Flows in Networks}.
\newblock Princeton University Press, 1962.

\bibitem{knuth}
D.~E. Knuth.
\newblock Two-way rounding.
\newblock {\em SIAM J. Discrete Math.}, 8:281--290, 1995.

\bibitem{monden2}
Y.~Monden.
\newblock What makes the {Toyota} production system really tick?
\newblock {\em Industrial Eng.}, 13:36--46, 1981.

\bibitem{monden1}
Y.~Monden.
\newblock {\em {Toyota} Production System}.
\newblock Industrial Engineering and Management Press, Norcross, GA, 1983.

\bibitem{ragh}
P.~Raghavan.
\newblock Probabilistic construction of deterministic algorithms: Approximating
  packing integer programs.
\newblock {\em J.~Comput. Syst. Sci.}, 37:130--143, 1988.

\bibitem{halficalp}
K.~Sadakane, N.~Takki{-C}hebihi, and T.~Tokuyama.
\newblock Combinatorics and algorithms on low-discrepancy roundings of a real
  sequence.
\newblock In {\em ICALP 2001}, volume 2076 of {\em Lecture Notes in Computer
  Science}, pages 166--177, Berlin Heidelberg, 2001. Springer-Verlag.

\bibitem{halfdagstuhl}
K.~Sadakane, N.~Takki{-C}hebihi, and T.~Tokuyama.
\newblock Discrepancy-based digital halftoning: Automatic evaluation and
  optimization.
\newblock In {\em Geometry, Morphology, and Computational Imaging}, volume 2616
  of {\em Lecture Notes in Computer Science}, pages 301--319, Berlin
  Heidelberg, 2003. Springer-Verlag.

\bibitem{10l}
J.~Spencer.
\newblock {\em Ten lectures on the probabilistic method}, volume~64 of {\em
  CBMS-NSF Regional Conference Series in Applied Mathematics}.
\newblock Society for Industrial and Applied Mathematics (SIAM), Philadelphia,
  PA, 1994.

\bibitem{steineryeomans}
G.~Steiner and S.~Yeomans.
\newblock Level schedules for mixed-model, just-in-time processes.
\newblock {\em Management Science}, 39:728--735, 1993.

\bibitem{lns155}
L.~Willenborg and T.~de~Waal.
\newblock {\em Elements of Statistical Disclosure Control}, volume 155 of {\em
  Lecture Notes in Statistics}.
\newblock Springer, 2001.

\end{thebibliography}
